\documentclass[twocolumn,secnumarabic,amssymb, amsmath, nofootinbib,tightenlines,
nobibnotes, aps, prl,epsfig]{revtex4}
\usepackage{graphicx}
\usepackage{dcolumn}
\usepackage{bm}
\usepackage{amssymb}
\usepackage{epsfig}
\usepackage{color}

\begin{document}

\newcommand{\ba}{\begin{eqnarray}}
\newcommand{\ea}{\end{eqnarray}}
\newcommand{\be}{\begin{equation}}
\newcommand{\ee}{\end{equation}}
\newcommand{\bdisplay}{\begin{displaymath}}
\newcommand{\edisplay}{\end{displaymath}}

\newcommand{\lldots}{,\dots,}
\newcommand{\ie}{\em i.e.,\  }
\newcommand{\lessabout}{\,\raisebox{-.6ex}{\ $\stackrel{<}{\sim }$\ }\,\,}
\newcommand{\greaterabout}{\,\raisebox{-.6ex}{\ $\stackrel{>}{\sim }$\ }\,}

\newcommand{\eq}[1]{Eq.\,(\ref{#1})}
\newcommand{\fig}[1]{Fig.\,\ref{#1}}

\newcommand{\la}{\,\raisebox{-.8ex}{\,$\stackrel{\textstyle <}{\sim}$}\,\,}
\newcommand{\ga}{\,\raisebox{-.8ex}{\,$\stackrel{\textstyle >}{\sim}$}\,\,}

\title{Heavy quark distributions from \\
the Color Dipole Picture}

\author{G.R. Boroun}
\email[]{boroun@razi.ac.ir}
\affiliation{Department of Physics, Razi University,\\
Kermanshah 67149, Iran}



\date{\today}

\begin{abstract}

We investigate charm-quark pair production at small Bjorken-$x$,
$x\leq 10^{-2}$, by combining the gluon distribution obtained from
the color dipole picture with the collinear generalized double
asymptotic scaling framework. The resulting reduced cross sections,
$\sigma_{\mathrm{red}}^{c\overline{c}}(W^2,Q^2)$, are found to be in
good agreement with the latest HERA measurements over a broad
kinematic range in $x$ and $Q^2$. The analysis yields an effective
pomeron intercept, while the color dipole model with the hard-pomeron
coefficient $C_{2}=0.29$ gives a comparable description in the
very-small-$x$ region, $x<10^{-3}$. Furthermore, the HERA data for
$2.5\leq Q^2\leq2000~\mathrm{GeV}^2$ exhibit the symmetry of the
saturation and color-transparency regimes when expressed in terms of
the scaling variable $\eta$. Including the threshold mass associated
with $J/\psi$ production in the color dipole picture shifts this
behavior toward the color-transparency region.\\
\end{abstract}

\pacs{}

\maketitle


\section{\label{Introduction}I. Introduction}
\renewcommand{\theequation}{\arabic{section}.\arabic{equation}}
\setcounter{section}{1}\setcounter{equation}{0}

HERA measurements of heavy-flavor production have provided valuable information on open charm and beauty production in neutral-current deep-inelastic electron-proton scattering. In particular, the combined H1 and ZEUS results \cite{HZ}, obtained by accounting for the correlations among statistical and systematic uncertainties, provide a comprehensive dataset for the reduced charm and beauty cross sections.

Within neutral-current deep-inelastic scattering, heavy-quark production is predominantly associated with the photon-gluon fusion mechanism,
$g\rightarrow c\overline{c}(b\overline{b})$, where $c$ and $b$ denote the charm and beauty quarks, respectively. The corresponding cross sections are sensitive to both the heavy-quark masses
 and the gluonic content of the proton. Their theoretical description therefore involves a wide range of perturbative scales $\mu^{2}$ in perturbative Quantum Chromodynamics (QCD). 
Two commonly employed frameworks are the massive fixed-flavour-number scheme (FFNS) \cite{FFNS} and the variable-flavour-number scheme (VFNS) \cite{VFNS}. The former is particularly appropriate
 in the vicinity of the heavy-quark threshold, $Q^{2}\approx m_{c,b}^{2}$, whereas the latter becomes more suitable when $Q^{2}\gg m_{c,b}^{2}$, where the resummation of collinear logarithms $\ln(Q^{2}/m_{c,b}^{2})$ is required. A general-mass variable-flavour-number scheme (GM-VFNS) has also been proposed for the treatment of heavy-quark contributions \cite{GFNS}.

Nikolaev and Zoller \cite{Zoller1} investigated the charm and beauty structure functions of the proton, \(F_2^c(x,Q^2)\) and \(F_2^b(x,Q^2)\), at small \(x\) within the color dipole approach and the BFKL-Regge framework\footnote{The corresponding coefficient functions are given in Ref.~\cite{Zoller2}.}. They demonstrated that, for the charm structure function at moderate \(Q^2\), the subleading BFKL contributions are strongly suppressed due to dynamical cancellations, resulting in the dominance of the rightmost hard BFKL pole with an intercept of \(\Delta_{\rm IP}\simeq 0.4\). In contrast, the beauty structure function receives significant contributions from subleading BFKL exchanges, leading to a different effective energy dependence.

Several theoretical approaches have been applied to heavy-quark production, including descriptions based on transverse-momentum-dependent (TMD) parton distributions in the Kimber-Martin-Ryskin (KMR) prescription \cite{KMR}. In this framework, the TMD distributions can be constructed from the standard parton distribution functions obtained within the generalized double asymptotic scaling (DAS) formalism. The corresponding high-energy behavior of the collinear coefficient functions for heavy-quark production has been investigated in Refs.\cite{Kotikov1, Kotikov2, Kotikov3, Kotikov4, Kotikov5}. Here, we study the combined H1 and ZEUS measurements \cite{HZ} of the reduced charm and beauty cross sections by using the generalized DAS approach together with analytical forms of the gluon density obtained within the color dipole picture.

At small values of Bjorken $x$, the color dipole picture (CDP) provides a useful representation of deep-inelastic scattering, following the original formulations of Refs.\cite{Sakurai, Nikolaev}. In this description, the virtual photon is treated through its fluctuation into a $q\overline{q}$ state, and the resulting dipole subsequently interacts with the proton. The photon fluctuation is characterized by a light-cone wave function, while the short-distance dynamics are described by perturbatively calculable QCD coefficient functions. The interaction corresponding to $\gamma^{*}g\to q\overline{q}$ can consequently be represented in terms of the exchange of gluons between the dipole and the gluonic field of the proton. The resulting formulation can be written in the factorized form
\begin{eqnarray} \label{eq:1.1}
\sigma_{L,T}^{\gamma^{*}p}(x,Q^{2})&=&\int dz
d^{2}\mathbf{r}_{\bot}
|\Psi_{\gamma}^{L,T}(\mathbf{r}_{\bot},z(1-z),Q^{2})|^{2}\nonumber\\
&&{\times}\widehat{\sigma}_{q\overline{q}}(\mathbf{r}_{\bot},W^{2}),
\end{eqnarray}
where $\widehat{\sigma}{q\overline{q}}(\mathbf{r}{\bot},W^{2})$ represents the interaction cross section of the color dipole with the proton. It is expressed in transverse-momentum space as
\begin{eqnarray}\label{eq:1.2}
\widehat{\sigma}_{(q\overline{q})p}(\mathbf{r}_{\bot},W^{2})&=&\int
d^{2}{\overrightarrow{\ell}}_{\bot}
\widetilde{\sigma}_{(q\overline{q})p}({\overrightarrow{\ell}}_{\bot}^{2},W^{2})\nonumber\\
&&{\times}(1-e^{-i{\overrightarrow{\ell}}_{\bot}{\overrightarrow{r}}_{\bot}}).
\end{eqnarray}
Here $\mathbf{r}{\bot}$ denotes the transverse distance between the quark and antiquark, whereas ${\overrightarrow{\ell}}{\bot}$ is the transverse momentum carried by the exchanged gluon. The first factor in the integrand is related to the transverse gluon-momentum distribution, while the second factor reflects the gauge structure of the QCD interaction \cite{Dieter}. The transverse-momentum representation of the dipole can be obtained by introducing $\overrightarrow{k_{\bot}}$ into the quark and antiquark four-momenta. Choosing the direction of $\overrightarrow{q}=\overrightarrow{k}+\overrightarrow{k}'$ as the $z$ axis gives
\begin{eqnarray}\label{eq:1.3}
\overrightarrow{k}=z\overrightarrow{q}+{\overrightarrow{k_{\bot}}},\nonumber\\
\overrightarrow{k}'=(1-z)\overrightarrow{q}-{\overrightarrow{k_{\bot}}}
\end{eqnarray}
with $\overrightarrow{k_{\bot}}.\overrightarrow{q}=0$. Within the CDP, $r(\equiv|\mathbf{r}|)$ specifies the transverse size of the $q\overline{q}$ configuration, while $z$ is the fraction of the incoming photon light-cone energy carried by the quark, with $0<z<1$. The corresponding photon Fock-state probability is described by $|\Psi_{\gamma}^{L,T}|^{2}$. The invariant mass of the dipole is related to the transverse momentum through$M^{2}_{q\overline{q}}=\frac{\overrightarrow{k_{\bot}}^{2}}{z(1-z)}$. For a center-of-mass energy $W$, the relevant $q\overline{q}$ configurations satisfy
$\frac{M^{2}_{q\overline{q}}}{W^{2}}\ll0.1$, while the Bjorken variable is approximately given by
$x{\cong}\frac{Q^{2}}{W^{2}}{\ll}0.1$.\

The interaction of the $q\overline{q}$ configuration with the gluonic field of the proton determines the behavior of the dipole cross section in the small-$r$ and large-$r$ domains. In the small-$r$ limit, the dipole undergoes color transparency, in accordance with the perturbative QCD description \cite{Bartels}. The gluon distribution associated with this region can be related to the transverse-momentum representation of the dipole cross section according to \cite{Dieter}
\begin{eqnarray}\label{eq:1.4}
\alpha_{s}(Q^{2})xg(x,Q^{2})=\frac{3}{4\pi}\int
d{\overrightarrow{\ell}}_{\bot}^{2}{\overrightarrow{\ell}}_{\bot}^{2}
\widetilde{\sigma}_{(q\overline{q})p}({\overrightarrow{\ell}}_{\bot}^{2},W^{2}).~~
\end{eqnarray}
The dependence of the total photoabsorption cross section on the scaling variable
$\eta(W^{2},Q^{2})=\frac{Q^{2}+m^{2}_{0}}{\Lambda^{2}_{sat}(W^{2})}$
was investigated experimentally in Ref.\cite{BKS}. The observed behavior can be represented by
\begin{equation}\label{eq:1.5}
\sigma^{\gamma^{*}p}~{\sim}~\sigma^{(\infty)}
 \begin{cases}
\frac{1}{\eta(W^{2},Q^{2})},~~~~~~~ \mathrm{for} ~\eta\gg1\\
\ln\frac{1}{\eta(W^{2},Q^{2})},~~~~ \mathrm{for} ~\eta\ll1.\\
 \end{cases}
\end{equation}
In this expression, $\sigma^{(\infty)}$ does not depend on the photon energy, $\Lambda_{sat}(W^{2})$ denotes the saturation scale, and $m_{0}^{2}\simeq0.15~\mathrm{GeV}^{2}$. The massive $q\overline{q}$ continuum consequently begins at a mass squared satisfying $m^2_0\lesssim m^2_{\rho^{0}}$, with $m^2_{\rho^0}$ representing the squared mass of the $\rho^0$ meson. The continuum of $q\overline{q}$ states is therefore described as a function of their invariant masses and provides a smooth extension of the low-lying vector-meson contributions.

For heavy-flavor production, the same scaling variable can be used after replacing $m_0^2$ by $m_{J/\Psi}^{2}$ for charm production and by $m_{\Upsilon}^{2}$ for beauty production. This prescription allows the scaling behavior of the $c\overline{c}$ and $b\overline{b}$ systems to be examined in terms of $\eta$. The quantity $\Lambda_{sat}^{2}(W^{2})$ determines the characteristic scale separating the two regimes. For $Q^{2}\gg\Lambda_{sat}^{2}(W^{2})$, the dipole interaction belongs to the color-transparency domain, whereas $Q^{2}\ll\Lambda_{sat}^{2}(W^{2})$ corresponds to the saturation domain. Thus, the condition $\eta(W^{2},Q^{2})=1$ provides the boundary between these two regions in the $(Q^{2},W^{2})$ plane. The domains $\eta(W^{2},Q^{2})<1$ and $\eta(W^{2},Q^{2})>1$ correspond, respectively, to saturation and color transparency.

At low $x$, the total photoabsorption cross section can consequently be regarded as a function of $\eta$ alone,
$\sigma_{\gamma^{}p}(W^{2},Q^{2})=\sigma_{\gamma^{}p}(\eta(W^{2},Q^{2}))$. Its limiting behavior is proportional to $\log(1/\eta(W^{2},Q^{2}))$ in the saturation region, $\eta(W^{2},Q^{2})<1$, and to $1/\eta(W^{2},Q^{2})$ for $\eta(W^{2},Q^{2})\gg1$. In the color-transparency domain, all interaction channels remain accessible and destructive interference is significant. In contrast, when $\eta(W^{2},Q^{2})\ll1$, corresponding to saturation, one of the channels becomes closed and the destructive interference is absent.\

Following Ref.\cite{Dieter}, the saturation scale is specified by
\begin{eqnarray}\label{eq:1.6}
\Lambda^{2}_{sat}(W^{2})=\frac{\pi}{\sigma^{(\infty)}}\int
d{\overrightarrow{\ell'}}_{\bot}^{2}{\overrightarrow{\ell'}}_{\bot}^{2}
\widetilde{\sigma}_{(q\overline{q})_{L}^{J=1}}({\overrightarrow{\ell'}}_{\bot}^{2},W^{2}),
\end{eqnarray}
where the dipole is considered in the fixed-spin state $J=1$, and $\ell'$ denotes the transverse momentum variable associated with the gluon. The relation between $\ell'$ and the transverse gluon momentum $\ell$ follows from the light-cone variable $z$,
\begin{eqnarray}\label{eq:1.7}
{\overrightarrow{\ell'}}_{\bot}^{2}=\frac{{\overrightarrow{\ell}}_{\bot}^{2}}{z(1-z)}.
\end{eqnarray}
The gluon density can then be connected directly to the saturation scale through
\begin{eqnarray}\label{eq:1.8}
\alpha_{s}(Q^{2})xg(x,Q^{2})=\frac{1}{8\pi}\sigma^{(\infty)}\Lambda^{2}_{sat}(W^{2}).
\end{eqnarray}
In this work, we investigate the dependence of the charm and beauty reduced cross sections on the gluon density at low and moderate values of $Q^{2}$ through their contribution to the DIS structure functions. The following section introduces the collinear generalized double asymptotic scaling framework, with the gluon density specified by the color dipole picture. Section III is devoted to the numerical analysis and the presentation of the obtained results. Finally, the calculated results are compared with the H1 and ZEUS measurements, followed by a summary of the main conclusions and remarks.\\

\section{II. Method}
\renewcommand{\theequation}{\arabic{section}.\arabic{equation}}
\setcounter{section}{2} \setcounter{equation}{0}

The heavy-quark structure functions measured in deep-inelastic electron-proton scattering can be obtained from the corresponding inclusive heavy-quark production cross sections. The experimentally measured differential cross section is conveniently written in terms of the reduced heavy-quark cross section, $\sigma_{\mathrm{red}}^{Q\overline{Q}}$, according to
\begin{eqnarray}\label{eq:2.1}
\frac{d^{2}\sigma^{{Q}\overline{{Q}}}}{dxdy}= \frac{2\pi
\alpha_{EM}^{2}}{xQ^{4}}\left(1-y+\frac{y^2}{2}\right)
\sigma_{\mathrm{red}}^{{Q}\overline{{Q}}}(x,Q^{2}).
\end{eqnarray}
The reduced quantity is related to the heavy-quark contributions to the transverse and longitudinal structure functions through
\begin{eqnarray}\label{eq:2.2}
\sigma_{\mathrm{red}}^{{Q}\overline{{Q}}}(x,Q^{2})=F_{2}^{Q}(x,Q^{2})-f(y)F_{L}^{Q}(x,Q^{2}),
\end{eqnarray}
where $f(y)=\frac{y^{2}}{1+(1-y)^{2}}$. Measurements performed at HERA\footnote{Diffractive production of high-mass states increases the cross section with increasing energy at fixed $Q^2$ and low $x$ \cite{Dieter}.} at small values of the Bjorken variable, $x{\cong}\frac{Q^{2}}{W^{2}}$, indicate an empirical scaling behavior of the reduced cross section. In particular, its dependence on $Q^{2}$ and $W$ can be represented in terms of a single scaling variable,
\begin{eqnarray}\label{eq:2.3}
\sigma_{\mathrm{red}}^{{Q}\overline{{Q}}}(W^2,Q^{2})=\sigma_{\mathrm{red}}^{{Q}\overline{{Q}}}(\eta_{Q\overline{Q}}),
\end{eqnarray}
which is consistent with the generalized vector-dominance/color-dipole picture (GVD/CDP) \cite{Cvetic}. At the kinematic limit specified below\footnote{In detail, we can express $x$ as
$$x=\frac{Q^2}{W^2+Q^2-M_{p}^2}{\cong}\frac{Q^2}{W^2}
$$
and $y$ denotes the ratio of the hadronic center-of-mass (COM) energy squared, $W^2$, to the total $e^{\pm}p$ energy squared, $s$, as
$$y=\frac{Q^2}{sx}{\cong}\frac{W^2}{s}.$$}
the longitudinal contribution of the virtual photon vanishes at $y=1$. Consequently, one obtains \cite{Boroun1}
\begin{eqnarray}\label{eq:2.4}
\sigma_{\mathrm{red}}^{{Q}\overline{{Q}}}(\eta_{Q\overline{Q}}){\simeq}F_{2}^{Q}(\eta_{Q\overline{Q}})-F_{L}^{Q}(\eta_{Q\overline{Q}}),
\end{eqnarray}
with
\begin{eqnarray}\label{eq:2.4n}
\eta_{Q\overline{Q}}=\frac{Q^{2}+m_{Q\overline{Q}}^{2}}{\Lambda^{2}_{sat}(W^{2})}.
\end{eqnarray}
At sufficiently small $x$, the gluon contribution provides the dominant component of the heavy-quark structure functions. Within the collinear
 generalized DAS framework, these functions can therefore be written as \cite{Kotikov5}
\begin{eqnarray} \label{eq:2.5}
F_{k=2,L}^{Q}(x,\mu^{2}){\simeq}~e^{2}_{Q}\sum_{n=0}(\frac{\alpha_{s}}{4\pi})^{n+1}B^{(n)}_{k,g}(x,\xi)
{\otimes} xg(x,\mu^{2}),~~
\end{eqnarray}
where $B_{k,g}$ denote the collinear Wilson coefficient functions in the high-energy limit, while $e^{2}_{Q}$ is the squared electric charge of the heavy quark. 
The index $n$ specifies the perturbative order in the running coupling $\alpha_s$, and $\xi=m_{Q}^{2}/\mu^{2}$. We take the renormalization and factorization scales to be identical, with the two choices
$\mu^{2}=Q^{2}+4m_{Q}^{2}$ and $\mu^{2}=Q^{2}$.\

The low-$x$ gluon distribution has also been investigated within the CDP framework in Ref.\cite{BKS} and subsequently employed in \cite{Machado}. It can be related to the proton structure function $F_{2}$ together with the ratio of longitudinal to transverse contributions. Defining $F_{2}=F_{T}+F_{L}$, this ratio is expressed through the corresponding photoabsorption cross sections as
\begin{eqnarray} \label{eq:2.6}
     R(W^2,Q^2) &{\equiv}&\frac{F_{L}(W^2,Q^2)}{F_{T}(W^2,Q^2)}=
\frac{\sigma_{\gamma^*_{L} p} (W^2,Q^2)}{\sigma_{\gamma^*_{T} p}
(W^2,Q^2)}.  \end{eqnarray}
Electromagnetic gauge invariance requires the longitudinal-to-transverse ratio to vanish in the real-photon limit,
 \be \label{eq:2.7}
      R(W^2,Q^2 = 0) = 0,
\ee 
whereas in the range relevant for the subsequent analysis,
$1<\eta(W^2,Q^2)<130$, the ratio can be approximated by
 \be \label{eq:2.8}
        R (W^2,Q^2) \approx \frac{1}{2 \rho},
 \ee
with $\rho$ taken to be approximately constant and close to unity. Thus, $\rho$ characterizes the relative longitudinal and transverse contributions to the photoabsorption cross section. For $\eta(W^2,Q^2)\gg\mu(W^2)$, one approximately has $R(W^2,Q^2)\simeq1/(2\rho)$, whereas the gauge-invariance condition gives $R(W^2,Q^2)=0$ at $Q^2=0$. For realistic values of $\rho$ near $\rho\cong1$, the total cross section is only weakly sensitive to this parameter. In the QCD evaluation of Ref.\cite{BKS}, the value $\rho=4/3$ is adopted, corresponding to the enhancement of the transverse $q\bar q$ fluctuation size relative to the longitudinal one.\

The transverse and longitudinal proton structure functions are connected to the corresponding virtual-photon absorption cross sections through
 \be \label{eq:2.9}
      F_{L,T} (W^2, Q^2) = \frac{Q^2}{4 \pi^2 \alpha}
        \sigma_{\gamma^*_{L,T} p} (W^2, Q^2),
\ee
and hence
\begin{eqnarray} \label{eq:2.10}
& F_2 (W^2, Q^2) = \frac{Q^2}{4 \pi^2 \alpha} \sigma_{\gamma^*p} (W^2, Q^2) \nonumber \\
& = \frac{Q^2}{4 \pi^2 \alpha} \left( \sigma_{\gamma^*_T p} (W^2,
Q^2) + \sigma_{\gamma^*_L p} (W^2, Q^2)\right).
\end{eqnarray}
Using the definition of $R(W^2,Q^2)$, the longitudinal structure function follows directly from $F_2$:
 \be \label{eq:2.11}
     F_L (W^2,Q^2) = \frac{R}{1+R} F_2 (W^2,Q^2).
 \ee
For the range $1<\eta(W^2,Q^2)<130$, this relation becomes
\begin{eqnarray} \label{eq:2.12}
F_L (W^2,Q^2) && = \frac{1}{2 \rho + 1} F_2 (W^2,Q^2),
\end{eqnarray}
In the color-transparency region, $Q^2\gg\Lambda^2_{sat}(W^2)$, the proton structure function can be described by the following two-parameter fit to the experimental data of Ref.\cite{Dieter}:
\begin{eqnarray} \label{eq:2.13}
       F_2 (W^2,Q^2) &=& \frac{R_{e^+e^-} \sigma^{(\infty)} (W^2)}{24 \pi^3}
       \frac{1+2 \rho}{3} \Lambda^2_{sat} (W^2) \nonumber \\
      & \times & \left(1 + 0 \left( \frac{1}{\eta} \right)
      \right),
\end{eqnarray}
where $R_{e^+e^-}=3\sum_qQ_q^2$, with the sum extending over the active quark flavors and $Q_q$ denoting their electric charges. Taking the saturation scale in the form
$\Lambda^2_{sat} (W^2) = C_1 \left( \frac{W^2}{1 {\rm GeV}^2}
\right)^{C_2}$
with $C_1=0.31~\mathrm{GeV}^2$ and $C_2=0.29$ (for further discussion, see the Appendix), Eq.~(\ref{eq:2.13}) can be recast as
\begin{eqnarray} \label{eq:2.14}
      F_2 (W^2,Q^2) && = \frac{R_{e^+e^-} \sigma^{(\infty)} (W^2)}{24 \pi^3}
   \frac{1+2 \rho}{3} C_1 \left( \frac{W^2}{1 {\rm GeV}^2} \right)^{C_2} \nonumber \\
    && \times \left( 1 + 0 \left( \frac{1}{\eta} \right) \right)
 \\
     && \equiv \hat{f}_2 (W^2) \left( \frac{W^2}{1 {\rm GeV}^2}
      \right)^{C_2} \left( 1 + 0 \left( \frac{1}{\eta} \right) \right),   \nonumber
\end{eqnarray}
The product $R_{e^+e^-}\sigma^{(\infty)}(W^2)$ is determined from the photoproduction cross section according to
 \be \label{eq:2.15}
         R_{e^+e^-} \sigma^{(\infty)} (W^2)
          = \frac{3 \pi}{\alpha} \frac{\sigma_{\gamma p} (W^2)}{\ln
          \frac{\rho \Lambda^2_{sat} (W^2)}{m^2_0}},
\ee
 where $\sigma_{\gamma p}(W^2)$ is obtained from a numerical evaluation followed by a $\left(\ln W^2\right)^2$ fit to the photoproduction data compiled by the Particle Data Group \cite{PDG}:
\begin{eqnarray}\label{eq:2.16}
\sigma_{\gamma p} (W^2) = & 0.003056 \left( 34.71 +
\frac{0.3894\pi}{M^2} \ln^2 \frac{W^2}{(M^2_p + M)^2}
\right) \nonumber \\
& + 0.0128 \left( \frac{(M_p + M)^2}{W^2} \right)^{0.462},
 \end{eqnarray}
where $M_p$ is the proton mass, $M=2.15~{\rm GeV}$, and $\sigma_{\gamma p}(W^2)$ is given in millibarns.\

The low-$x$ gluon distribution can consequently be written in terms of the structure function as \cite{BKS}
\begin{eqnarray} \label{eq:2.17}
      && xg(x,Q^2) = \frac{9 \pi}{\alpha_s (Q^2) R_{e^+e^-}} \frac{1}{(2 \rho + 1)}
             F_2 (\xi_L x, Q^2) \nonumber \\
      &=& \frac{9 \pi}{R_{e^+e^-}} \frac{1}{(2 \rho + 1)} \hat f_2 \xi_L^{-C_{2}} \left(
          \frac{W^2}{1 {\rm GeV}^2} \right)^{C_2},
\end{eqnarray}
 where
\begin{eqnarray} \label{eq:2.18}
\hat f_2 = \frac{C_{1}(1+2\rho)}{24 \pi^2\alpha}
\frac{\sigma_{\gamma p} (W^2)}{\ln
          \frac{\rho \Lambda^2_{sat} (W^2)}{m_{Q\overline{Q}}^{2}}},
\end{eqnarray}
with $\xi_L=0.4$ and $\rho=\frac{4}{3}$. Equation~(\ref{eq:2.17}) corresponds to the asymptotic form of the complete calculation presented in Ref.\cite{BKS}. At the initial scale $Q_0^2=1.9~\mathrm{GeV}^2$, it can be parameterized as
$xg(x,Q_0^2)\approx0.5x^{-0.21}(1-x)^6$, which is compatible with the results given in Table III of Ref.\cite{Machado}.\

The reduced cross section for heavy-quark pair production can now be obtained by combining the above gluon density with the heavy-quark structure functions in the CDP framework:
\begin{eqnarray}\label{eq:2.19}
\sigma_{\mathrm{red}}^{{Q}\overline{{Q}}}(W^2,Q^2){\simeq}\frac{9
{\pi}e^{2}_{Q}}{R_{e^+e^-}} \frac{1}{(2 \rho + 1)} \hat f_2
\xi_L^{-C_2}\sum_{n=0}(\frac{\alpha_{s}}{4\pi})^{n+1}{\times}\nonumber\\
\left[B^{(n)}_{2,g}(W^{2},\xi)-f(y)B^{(n)}_{L,g}(W^{2},\xi)
\right] {\otimes} \left(\frac{W^2}{1 {\rm GeV}^2}
\right)^{C_2}.~~~
\end{eqnarray}
In the following section, the CDP gluon density is incorporated into the collinear DAS formalism to calculate the reduced cross sections for charm-pair production in the low-$x$ region.\\

\section{III. The result and Conclusion}
\renewcommand{\theequation}{\arabic{section}.\arabic{equation}}
\setcounter{section}{3}\setcounter{equation}{0}

We adopt the standard QCD running coupling, with its normalization
fixed at the $Z$-boson mass according to
$\alpha_{s}(M_{Z}^{2})=0.1166$. Since the heavy-quark mass enters the
virtual-photon wave function in the color dipole picture, the Bjorken
variable appearing in the dipole cross section is modified to account
for the production threshold. Accordingly, we make the replacement
\begin{eqnarray}\label{eq:3.2}
x{\rightarrow}x\left(1+\frac{4m_{Q}^{2}}{Q^2}\right),
\end{eqnarray}
where the running masses used for the charm and beauty quarks are
$m_{c}=1.29~\mathrm{GeV}$ and $m_{b}=4.049~\mathrm{GeV}$,
respectively \cite{HZ}. The boundary between the saturation and color
transparency domains is likewise affected by the heavy-quark
production threshold. In the scaling variable, this amounts to the
replacement $m_{0}^{2}\rightarrow m_{Q\overline{Q}}^{2}$, with
$
m_{J/\Psi}(1s)=3097\pm0.011~\mathrm{MeV}, \qquad
m_{\Upsilon}(1s)=9460\pm0.26~\mathrm{MeV},
$
as given in Ref.\cite{PDG1}.\

We first examine the reduced charm-production cross section measured
at HERA \cite{HZ} as a function of the scaling variable $\eta$. When
the light-quark scale $m_{0}^{2}$ in Eq.~(\ref{eq:2.4n}) is used, the
data exhibit contributions from both the saturation domain,
$\eta(W^{2},Q^{2})<1$, and the color-transparency domain,
$\eta(W^{2},Q^{2})>1$. Replacing this scale by the $J/\psi$ threshold,
$m_{0}^{2}\rightarrow m_{J/\Psi}^{2}$, moves the data predominantly
toward the color-transparency region, as displayed in Fig.~1. The
numerical analysis shown in the figure employs
$\rho=4/3$, $C_{2}=0.29$, and $\xi_{L}=0.4$.\

An additional feature becomes apparent when the HERA measurements are
represented using $\eta$ with $m_{0}^{2}=0.15~\mathrm{GeV}^{2}$. The
reduced charm cross section displays an approximately symmetric
behavior around $\eta\simeq8-9$. In particular, the large- and
small-$\eta$ regions exhibit an approximate invariance under the
transformation $\eta\leftrightarrow1/\eta$ over the considered range
\cite{Stasto}. This behavior changes once the scale $m_{0}^{2}$ is
replaced by $m_{J/\Psi}^{2}$, after which the measurements are shifted
toward the color-transparency side. The observed symmetry properties
are therefore sensitive to the coefficient entering the saturation
scale $\Lambda_{\mathrm{sat}}$. For this reason, we next express the
results directly in terms of the invariant energy $W^{2}$.\

\begin{figure}[h]

\includegraphics[width=9cm]{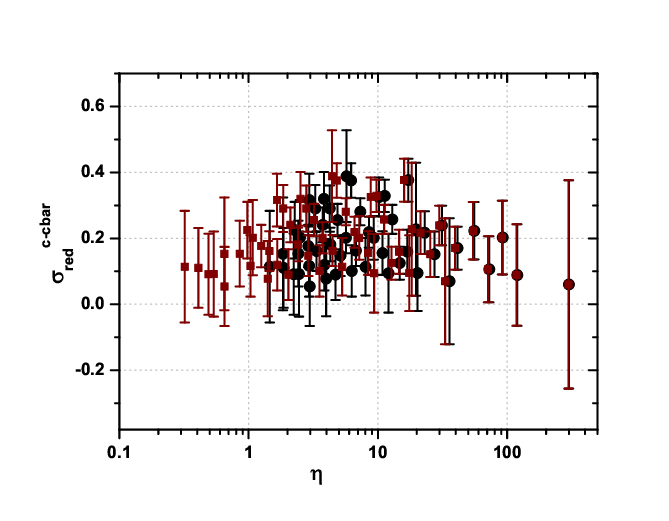}%

\caption{\label{Fig1}
The HERA measurements \cite{HZ} of
$\sigma_{\mathrm{red}}^{c\overline{c}}$ represented in terms of the
scaling variable $\eta$ defined in Eq.~(\ref{eq:2.4n}). The red-square
points correspond to $m_{0}^{2}=0.15~\mathrm{GeV}^{2}$, while the
black-circle points employ the $J/\psi$ threshold,
$m_{J/\Psi}^{2}=9.59~\mathrm{GeV}^{2}$. The shift from the saturation
and color-transparency regions toward the color-transparency domain is
shown. The experimental points include the total uncertainties.}

\end{figure}

We subsequently study the $W^{2}$ dependence of the reduced charm
cross section within the collinear DAS formulation combined with the
CDP. Figure~2 presents the calculation at the scale
$\mu^{2}=Q^{2}+4m_{c}^{2}$ for four representative values of $x$=
$10^{-5}$, $10^{-4}$, $10^{-3}$, and $10^{-2}$. These predictions are
compared with the combined H1 and ZEUS measurements \cite{HZ},
including their total uncertainties. The experimental data cover the
range $2.5\leq Q^{2}\leq2000~\mathrm{GeV}^{2}$ and correspond to
deep-inelastic $ep$ scattering at the center-of-mass energy
$\sqrt{s}=318~\mathrm{GeV}$.\

\begin{figure}[h]

\includegraphics[width=9cm]{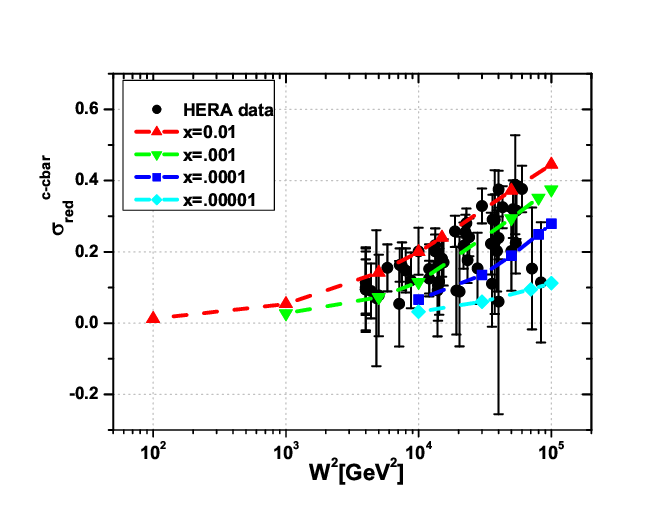}%

\caption{\label{Fig2}
The reduced charm-production cross section
$\sigma_{\mathrm{red}}^{c\overline{c}}$ as a function of $W^{2}$,
calculated at $\mu^{2}=Q^{2}+4m_{c}^{2}$ for
$x=10^{-5}$ (cyan-square curve), $x=10^{-4}$ (blue-square curve),
$x=10^{-3}$ (green-down-triangle curve), and
$x=10^{-2}$ (red-up-triangle curve). The theoretical results are
compared with the HERA data \cite{HZ}, including total errors.}

\end{figure}

At the two smallest values of $x$, namely $x=10^{-5}$ and $10^{-4}$,
we use the hard-pomeron coefficient $C_{2}=0.29$. For the larger
values $x=10^{-3}$ and $10^{-2}$, the comparison with the data favors
smaller effective values, approximately $C_{2}=0.24$ and $0.21$,
respectively. Thus, the exponent controlling the energy dependence
cannot be regarded as strictly universal over the complete kinematic
range. This behavior is consistent with the effective-pomeron
description of the proton structure function discussed in
Ref.\cite{Desgr}.\

The comparison also indicates that the effective coefficient
$C_{2}(x,Q^{2})$ associated with charm production has a non-negligible
dependence on the kinematic variables, particularly as $x$ increases.
Consequently, treating $C_{2}$ as an exactly constant quantity over a
finite $x$ interval is not justified. The variation observed in the
present analysis is obtained from the comparison with the HERA
measurements of $\sigma_{\mathrm{red}}^{c\overline{c}}$ in the range
$0<y\leq1$. Its $Q^{2}$ dependence is related to the gluon evolution
in the double-logarithmic limit, whereas the corresponding exponent
remains approximately fixed with increasing $Q^{2}$ in the GBW model
\cite{Golec}, as discussed in Ref.\cite{Beuf}.\

Figure~3 examines the $x$ dependence of the charm reduced cross
section at moderate virtualities. We consider
$Q^{2}=2.5$, $7$, and $18~\mathrm{GeV}^{2}$ and use the corresponding
effective pomeron coefficients. The predictions remain compatible
with the HERA measurements \cite{HZ} throughout the considered
$0<y\leq1$ interval. As $x$ becomes larger, the reduced cross section
decreases. Conversely, toward smaller $x$ and large inelasticity,
approaching $y=1$, the cross section tends toward a limiting behavior,
as indicated by the final points in Fig.~3.\

\begin{figure}[h]

\includegraphics[width=9cm]{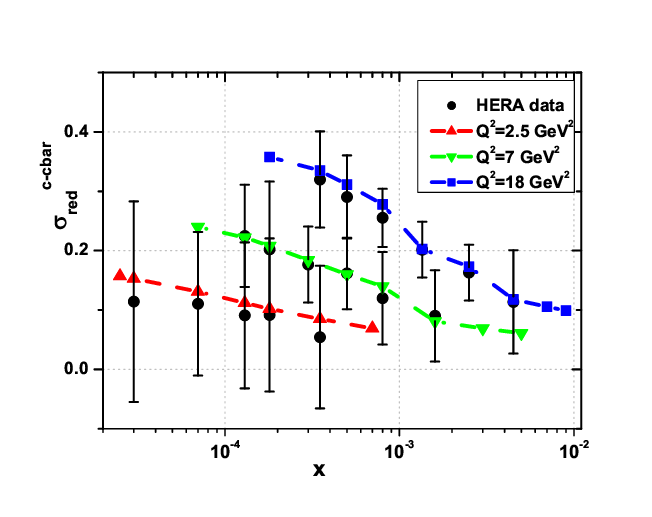}%

\caption{\label{Fig3}
The $x$ dependence of $\sigma_{\mathrm{red}}^{c\overline{c}}$
calculated at $\mu^{2}=Q^{2}+4m_{c}^{2}$ for
$Q^{2}=2.5~\mathrm{GeV}^{2}$ (red-up-triangle curve),
$Q^{2}=7~\mathrm{GeV}^{2}$ (green-down-triangle curve), and
$Q^{2}=18~\mathrm{GeV}^{2}$ (blue-square curve). The predictions
obtained with the effective pomeron contributions in the range
$0<y\leq1$ are compared with the HERA data \cite{HZ}, including total
uncertainties.}

\end{figure}

Finally, we focus on the very-small-$x$ domain, $x<10^{-3}$, where
the intercept is fixed at $C_{2}=0.29$. Figure~4 shows the resulting
$W^{2}$ dependence of $\sigma_{\mathrm{red}}^{c\overline{c}}$ within
the collinear DAS approach based on the CDP. The comparison with the
HERA measurements in the range
$2.5\leq Q^{2}\leq32~\mathrm{GeV}^{2}$ shows good agreement, including
the experimental uncertainties. These results indicate that the CDP
with $C_{2}=0.29$, as predicted in the literature, provides a suitable
description of the charm reduced cross section at very small $x$.
For increasing $x$ and $Q^{2}$, an effective pomeron coefficient is
required to maintain agreement with the measurements.\

\begin{figure}[h]

\includegraphics[width=9cm]{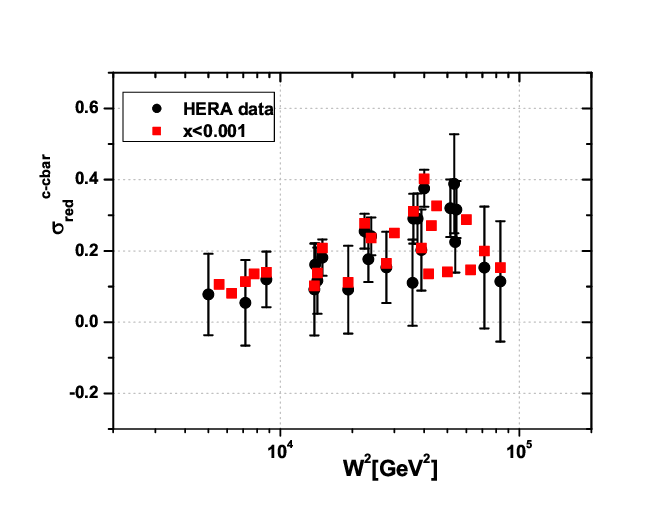}%

\caption{\label{Fig4}
The reduced charm-production cross section
$\sigma_{\mathrm{red}}^{c\overline{c}}$ as a function of $W^{2}$,
calculated at $\mu^{2}=Q^{2}+4m_{c}^{2}$ for $x<10^{-3}$ using the
fixed value $C_{2}=0.29$ (red-square data). The results are compared
with the HERA measurements \cite{HZ} in the range
$2.5\leq Q^{2}\leq32~\mathrm{GeV}^{2}$, including total uncertainties.}

\end{figure}

In conclusion, we have investigated charm-quark pair production at
small Bjorken-$x$ using the color dipole picture together with the
corresponding gluon distribution in the collinear generalized double
asymptotic scaling framework. The reduced cross section
$\sigma_{\mathrm{red}}^{c\overline{c}}(W^{2},Q^{2})$ has been
examined over a broad range of $W^{2}$ in order to determine the
effective pomeron coefficient governing its energy dependence. The
comparison with the HERA measurements shows that the CDP provides a
consistent description of the charm-production data, with the
hard-pomeron contribution playing an important role.\

The analysis further indicates that the effective value of $C_{2}$
decreases as $x$ and $Q^{2}$ increase. In the very-small-$x$ region,
the fixed value $C_{2}=0.29$ gives results compatible with the
experimental measurements. We have also investigated the
$\eta$-scaling behavior by incorporating the photoabsorption cross
section for $c\overline{c}$ production. The transition between the
saturation and color-transparency domains is modified when the
threshold mass of the $J/\psi$ state is used instead of $m_{0}^{2}$,
leading to a displacement of the data toward the color-transparency
region. Overall, the theoretical results obtained within the present
framework are in good agreement with the HERA data at small $x$,
with the longitudinal charm structure-function contribution
$F_{L}^{c\overline{c}}(W^{2},Q^{2})$ contributing significantly to
the description of the reduced cross section.\\


\section{DECLARATION OF AI-ASSISTED LANGUAGE EDITING}

The author has used AI-assisted tools solely for language editing and
improving the readability of the manuscript. The author has carefully
reviewed and revised the manuscript and takes full responsibility for
its scientific content.\\

\section{Appendix:}

A description of the proton structure function over a broad range of
$Q^2$, extending from the low- to the high-$Q^2$ region, requires
consideration of both soft- and hard-pomeron contributions. The
tensor-pomeron framework and its application to low-$x$ DIS and
photoproduction have been discussed in Ref.\cite{Ewerz}. Along with
the soft tensor pomeron, which successfully describes soft hadronic
interactions at high energies, a hard tensor-pomeron contribution can
also be introduced. Its applicability within the color dipole picture
has been investigated in Ref.\cite{Dieter1}. The two-tensor-pomeron
model of Ref.\cite{Cvetic} describes the transition between the
low-$Q^2$ domain, in which the real or virtual photon exhibits
hadron-like behavior, and the large-$Q^2$ regime characterized by
hard scattering.

The corresponding pomeron intercepts are specified by
$\alpha_{1}(0)=1+\epsilon_{1}$ for the soft contribution, with
$\epsilon_{1}=0.0935$, in agreement with the conventional value of
approximately $0.08$ \cite{DL0,DL1}. For the hard pomeron, the
intercept is given by $\alpha_{0}(0)=1+\epsilon_{0}$, where
$\epsilon_{0}=0.3008$. This value is also consistent with the results
reported in Refs.\cite{DL0,DL1}. Within this framework, the
photoabsorption cross sections for both real and virtual photons
scattering from the proton can be treated in a common description.\

The tensor-pomeron representation of the proton structure function was
studied in Ref.\cite{DL0}. The corresponding experimental data indicate
that, in addition to the soft-pomeron contribution, a hard-pomeron
term is required to describe high-energy hadronic processes. The
structure function can consequently be written in the form
\begin{eqnarray}\label{eq:2.19}
F_{2}(W^2,Q^2){\sim}
\sum_{i=0,1}f_{i}(Q^2)(W^2)^{\epsilon_{i}},
\end{eqnarray}
where the functions $f_i(Q^2)$ are parameterized in
Refs.\cite{DL1,DL2}. These parameterizations provide a good
description of the available small-$x$ measurements of $F_2$ together
with the photoproduction cross section $\sigma^{\gamma p}$.\

A further contribution is required when Regge factorization is applied
to the different processes. As discussed in Ref.\cite{DL2}, at a
fixed value of $W=\sqrt{s}$ corresponding to $y=1$, the relevant
contributions can be expressed as

\begin{eqnarray}\label{eq:2.20}
F_{2}(W^2,Q^2){\sim}
\sum_{i=0,1,R}f_{i}(Q^2)(W^2)^{\epsilon_{i}},
\end{eqnarray}
where the Reggeon exponent is $\epsilon_R=-0.476$. Thus, the Regge
factorization prescription can be extended to the charm-production
structure functions in addition to the soft- and hard-pomeron terms.
For the charm structure function, powers of $(1-x)$ are included in
the individual contributions so that the appropriate behavior is
obtained as $x\rightarrow1$.\\


\end{document}